# To Click or Not To Click:
# Automatic Selection of Beautiful Thumbnails from Videos

Yale Song, Miriam Redi*, Jordi Vallmitjana, Alejandro Jaimes*
Yahoo Research
{yalesong,redi,jvallmi,ajaimes}@yahoo-inc.com

Code and dataset: https://github.com/yahoo/hecate, https://github.com/yalesong/thumbnail

## ABSTRACT

Thumbnails play such an important role in online videos. As the most representative snapshot, they capture the essence of a video and provide the first impression to the viewers; ultimately, a great thumbnail makes a video more attractive to click and watch. We present an automatic thumbnail selection system that exploits two important characteristics commonly associated with meaningful and attractive thumbnails: high relevance to video content and superior visual aesthetic quality. Our system selects attractive thumbnails by analyzing various visual quality and aesthetic metrics of video frames, and performs a clustering analysis to determine the relevance to video content, thus making the resulting thumbnails more representative of the video. On the task of predicting thumbnails chosen by professional video editors, we demonstrate the effectiveness of our system against six baseline methods, using a real-world dataset of 1,118 videos collected from Yahoo Screen. In addition, we study *what makes a frame a good thumbnail* by analyzing the statistical relationship between thumbnail frames and non-thumbnail frames in terms of various image quality features. Our study suggests that the selection of a good thumbnail is highly correlated with objective visual quality metrics, such as the frame texture and sharpness, implying the possibility of building an automatic thumbnail selection system based on visual aesthetics.

## 1. INTRODUCTION

Thumbnails play such an important role in online videos. As the most representative snapshot, they capture the essence of videos and provide the first impression to the viewers. Studies suggest that people look at thumbnails extensively when browsing online videos [5], and often are a crucial decision factor in determining whether to watch a video or skip to another [8]. A great thumbnail ultimately makes a video more attractive to watch, which, in turn, leads to the increase in ad revenue [18]. Due to its importance, video producers and editors put an increasing amount of effort into selecting meaningful and attractive thumbnails.

*Redi and Jaimes performed work while working at Yahoo.

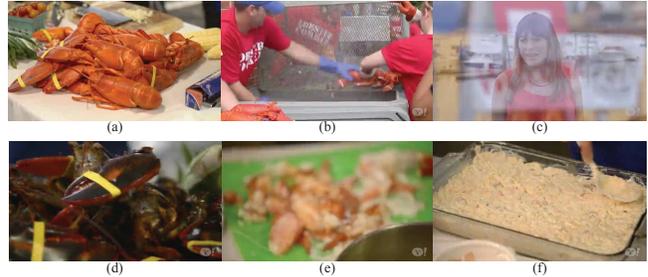

Figure 1: Among the six thumbnail candidates of a video *Maine Lobster Festival*, which one would you rather click to watch? Some images are less attractive than others, e.g., (b) has a poor composition, (c) is a frame in transition (dissolve), (d) is too dark, and (e) is out of focus; others are less relevant to the main topic, e.g., it is not immediately obvious why (c,f) are related to Lobster Festival. We present an automatic thumbnail selection system that exploits two important characteristics of good thumbnails, the relevance to video content and the aesthetic quality of frames, allowing us to obtain the most relevant and attractive thumbnail, e.g., (a) lobsters.

Unfortunately, manual selection of thumbnails can be cumbersome in many scenarios. For example, user generated videos, such as the ones uploaded to Flickr and Tumblr, generally do not come with thumbnails chosen by the users. Also, videos produced by third parties and delivered to media companies sometimes come without thumbnails. Selecting a thumbnail manually from even a few minutes of those videos can be extremely time consuming, making it difficult to scale. Our goal is to automate this process by coming up with a single best frame suitable as thumbnail, or by suggesting just a few frames for humans to choose from.

We present an automatic thumbnail selection system that exploits two important characteristics commonly associated with high quality video thumbnails: high relevance to video content and superior aesthetic quality.[1] We argue that high relevance alone is insufficient because people are naturally drawn to attractive images. Superior aesthetic quality alone is similarly insufficient because videos with attractive, but irrelevant thumbnails (e.g., clickbaits) may lead to user disappointment, which could ultimately result in negative reviews and harm the reputation of video providers [21].

Our system selects *attractive* thumbnails by analyzing various visual quality and aesthetic metrics of video frames, and performs a clustering analysis to determine the relevance to video content, thus making resulting thumbnails more *representative* of the video. While previous work has addressed one aspect or another (e.g., by selecting thematic [13], query-sensitive [23, 25], and interesting thumbnails [14]), the ability to consider both the relevance and the attractiveness for thumbnail selection appears, to the best of our knowledge, to be unique to our work.

---
[1] Another characteristic of good thumbnails is an appropriate inclusion of interesting "things" (e.g., face, object, text, etc.). We do not explicitly consider those high-level semantics in this work because it is heavily dependent on the types of video content.

The key benefit of our system is that it is computationally efficient. This makes it especially favorable in the production environment. Our system discards low quality and near duplicate frames early in the process so that we can focus on processing only the promising frames. Also, an unsupervised version of our approach does not rely on any machine learning technique besides clustering. As a result, our system produces a thumbnail in only under 10% of video length, on a conventional laptop computer using a single CPU. This low computational footprint and the simple design of the architecture mean our system can be further extended in the future to benefit from more sophisticated techniques.

Evaluation of automatic thumbnail selection is challenging due to the lack of ground-truth and the subjectivity inherent in the task. Previous studies have used a small number of videos [23], a limited number of baseline methods [13], and a rather too simple design of experiments [12]. In this paper, we evaluate our system on a task of predicting thumbnails chosen by professional video editors, using a dataset of 1,118 videos collected from Yahoo Screen.[2] The dataset contains high-quality thumbnails chosen manually by professional video editors. This allows us to measure quantitatively, and automatically, how well our method can predict thumbnails chosen by humans – a challenging real-world task.

In addition to the experiments, we study *what makes a frame a good thumbnail* by performing a statistical analysis of various visual features extracted from editorially chosen thumbnails. Our study suggests that the selection of good thumbnails is highly correlated with objective visual quality metrics, such as the frame texture and sharpness – implying the possibility of building an automatic thumbnail selection system based on objective quality metrics – but not so much with the standard photographic beauty rules, such as the Rule of Thirds.

In summary, we make the following contributions in this paper:
- We present an automatic thumbnail selection system that exploits the relevance to video content and the visual attractiveness of images. Our code is available at https://github.com/yahoo/hecate
- We collected a dataset of 1,118 videos and their associated thumbnails chosen by professional video editors. Our dataset is available at https://github.com/yalesong/thumbnail
- We introduce a framework for automatically evaluating thumbnail extraction techniques on online videos. On our dataset, we show that our technique significantly outperforms several existing baselines.
- Finally, we study *what makes a frame a good thumbnail* and report several findings that could help design an automatic thumbnail selection system.

## 2. RELATED WORK

Our work is closely related to three areas in multimedia understanding: thumbnail selection, video highlighting and summarization, and computational aesthetics. Below we review and differentiate our work from some of the most related works in those areas.

---
[2]As of January 2016, Yahoo Screen has been sunset; most video content is available on various Yahoo websites instead.

**Thumbnail selection:** Few works in the literature have specifically addressed automatic thumbnail selection. Gao *et al.* [13] proposed a thematic thumbnail selection framework, which chooses a thumbnail that has the most common visual characteristics shared with an extra set of web images, obtained using keywords from the video. Similarly, Liu *et al.* [23] proposed a query-specific thumbnail selection algorithm that extracts a frame that is both representative of video content and is specific to the intent of user query, by matching query words and frame content using visual concept detectors. Liu *et al.* [25] also proposed a query-specific thumbnail selection method by learning a joint embedding space of visual and textual signals.

While the previous work has approached thumbnail selection from various perspectives, little attention has been paid on the *attractiveness* of the thumbnails, i.e. their capability to attract the users' attention using visual quality and aesthetics criteria. Our method is explicitly designed to discover attractive thumbnails using several visual quality metrics and computational aesthetics.

Moreover, the performances of existing frameworks are evaluated either on a small number of videos [13, 23] or with a rather too simple design of experiments (e.g., compared against either no baseline [13] or only one baseline that selects the middle frame of a video [12]). In this paper, we evaluate our system against six baseline approaches on a set of 1,118 videos, the largest of its kind to the best of our knowledge. In addition, we also study what makes a frame a good thumbnail by investigating visual features commonly shared among thumbnails chosen by professional video editors.

**Video highlighting and summarization:** The goal of video highlighting is to find the most important segments from a video. Video summarization goes a step further and aims to generate the *gist* of a video by carefully selecting a few highlight frames or segments so that they deliver the video content in the most compact form [29]. While the two tasks have slightly different goals from thumbnail selection, there has been an overlap of techniques used.

Cong *et al.* [7] proposed a sparse dictionary selection approach, which aims to reconstruct a video sequence from only a few "basis frames" from the video using group LASSO. Zhao and Xing [46] have extended this approach into an online dictionary learning problem. Central to those work is content non-redundancy in the summary. Our work not only removes redundant frames, we also consider visual quality and aesthetics to select a single most beautiful thumbnail. Song *et al.* [37] used the title and description to identify important segments in a video; Chu *et al.* [6] used visual co-occurrence across multiple videos to measure visual interestingness; and Gygli *et al.* [15] used user-generated animated GIFs to identify video highlights. Our work is different from that line of work in that, while they identify important frames by measuring visual similarity to the reference (e.g., title [37]), we directly assess the photographic attractiveness to identity "beautiful" frames.

**Computational aesthetics:** Computational aesthetics aims to measure the photographic quality of images using visual analysis techniques. The first work in this field has appeared less than a decade ago [9, 19], which proposed to distinguish amateur vs. professional photographs based on visual features inspired by a photographic theory. Since then, computational aesthetics has mainly been applied to the task of aesthetic image classification [32, 11, 30] and ranking [41]. Other applications include video creativity assessment [33], video interestingness estimation [14], and machine-assisted image composition [43] and enhancement [3]. To the best of our knowledge, our work represents the first attempt to evaluate the visual aesthetic quality of video frames for automatic thumbnail selection.

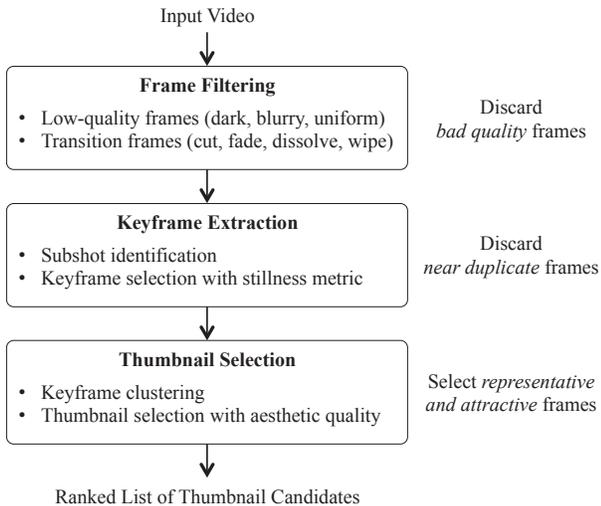

Figure 2: A schematic diagram of our automatic thumbnail selection system. On the right side are the goals of each step.

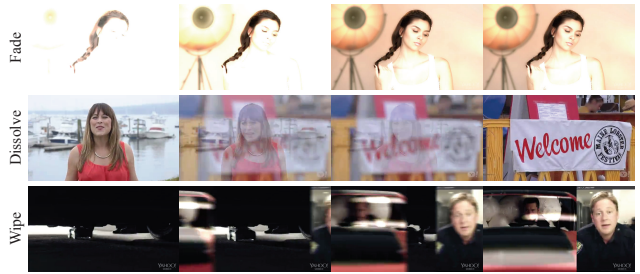

Figure 3: Frames during a shot transition are usually of low-quality. Shown here are examples of fade in, dissolve, and wipe. We perform shot boundary detection and filter out frames around each boundary, eliminating frames shown above.

## 3. OUR METHOD

Our system consists of three steps: frame filtering, keyframe extraction, and thumbnail selection. Figure 2 illustrates a schematic overview of our system. For the clarity of the paper, we define a glossary of the terms used in this section: *frames* refer to the entire frames of a video; *keyframes* refer to a compact and non-redundant subset of the video frames; and a *thumbnail* refer to the most representative and attractive frame of a video.

### 3.1 Frame Filtering

Processing the entire frames in a video can be quite time consuming; many video processing systems reduce the number of frames by subsampling at a uniform time interval. For the purpose of thumbnail selection, however, the subsampling method can accidentally discard frames ideal as a thumbnail, making it undesirable for this work. Intuitively, and as confirmed by our findings in Section 6, good thumbnails must possess certain aesthetic qualities, e.g., brightness, sharpness, colorfulness, etc. This allows us to reduce the number of frames by filtering out low-quality frames, improving processing efficiency in the later stage.

**Low-quality frames**: We detect and filter out three types of low-quality frames: dark, blurry, and uniform-colored. Specifically, we filter out dark frames by computing relative luminance [2] as

$$Luminance(I_{rgb}) = 0.2126 I_r + 0.7152 I_g + 0.0722 I_b \quad (1)$$

and thresholding it by an empirically chosen value. We filter out blurry images in a similar way by computing the sharpness as

$$Sharpness(I_{gray}) = \sqrt{(\Delta_x I_{gray})^2 + (\Delta_y I_{gray})^2} \quad (2)$$

Finally, to filter out uniform-colored frames, we first compute a normalized intensity histogram of an image, sort the values in a descending order, compute a cumulative distribution at top 5% bins, and threshold it with an empirically chosen value.

$$Uniformity(I_{gray}) = \int_0^{5\%} cdf(sort(hist(I_{gray}))) dp \quad (3)$$

**Transitioning frames:** We observe that frames during a shot transition (especially fade in/out, dissolve, and wipe) are usually of low-quality (see Figure 3); we filter them out by performing shot boundary detection. Extensive efforts have been made to shot boundary detection before [36]. For the simplicity and efficiency of the solution, we detect shot boundaries by using the method developed by Zabih *et al.* [44], which computes the edge change ratio between two consecutive frames and detects a boundary with thresholding. We then filter out frames around each boundary.

Our low-quality and transitioning frame filtering methods are simple to implement and fast to compute, while efficiently reducing the number of frames. In our dataset of 1,118 videos (see Section 4), the average processing time of the two steps together was only 0.3% relative to the video length (e.g., it takes 0.9 second for a 5 minute video), while reducing the total number of frames down to 88.64% on average.

### 3.2 Keyframe Extraction

A video sequence, by nature, has many near duplicate frames; we further improve the computational efficiency of our system by filtering out near duplicate frames via keyframe extraction.

There exists many techniques to extract keyframes. The most efficient, perhaps the most frequently used approach is based on clustering analysis. It clusters frames by their visual similarity and selects the most representative frames, one per cluster, by selecting a frame closest to the centroid (using the k-means algorithm [10]) or the medoid (using the k-medoids algorithm [26]) of samples within each cluster.

In this work, different from the conventional approaches, we use image aesthetics (especially, the stillness) to select a frame from each cluster. This is motivated by the well-known blurring artifact caused by motion compensation during video compression [1]. We observe that frames appear as blurred when there is high motion energy; said slightly differently, we can expect to obtain sharper, higher quality keyframes if we select frames that have low motion energy. This observation leads to a simple, yet quite powerful method for keyframe generation, described below.

**Feature extraction:** We start by obtaining feature representation from the remaining frames. Since our goal is to identify near duplicates, simple easy-to-compute image descriptors, such as color and edge histograms would suffice. We compute two types of features: a pyramid of HSV histograms with 128 bins per channel, and a pyramid of edge orientations and magnitudes with 30 bins for each. The features are computed over a two-level spatial pyramid (five regions), resulting in a 2,220 dimensional feature vector.

**Subshot identification:** It is better to over-sample than to under-sample the keyframes, because in the next step the thumbnails will be chosen only from the remaining keyframes. Extracting one keyframe per shot, a continuous block of the remaining frames found after the frame filtering step, may accidentally discard frames suitable as a thumbnail. We therefore identify subshots within each shot, and extract a keyframe from each subshot. We identify subshots by clustering the remaining frames using the k-means algorithm. We set the number of clusters to the number of shots in a video; we empirically found that this consistently leads to good clustering results. A subshot within a shot is then identified as a continuous block of frames with the same cluster ID.

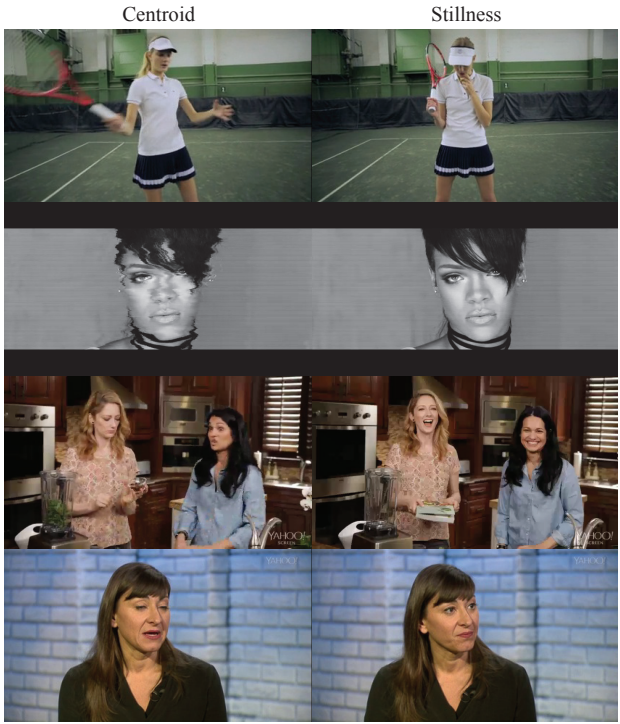

**Figure 4: Keyframes selected using the conventional approach (left) and our approach (right). Selecting a frame with the minimum frame difference value, thus the most *still* among its neighbor frames, allows us to remove redundant frames and to obtain sharper images desirable for a thumbnail.**

| | Feature | Dim | Description |
|---|---|---|---|
| **Color** | *Contrast* [34] | 1 | Ratio between luminance range and average luminance. |
| | *HSV* [27] | 3 | Average on each channel of Hue, Saturation, and Brightness. |
| | *HSV (Central)* [27] | 3 | Average HSV computed on the central quadrant of the image. |
| | *HSV Histograms* [27] | 20 | Histograms of HSV channels with 12, 3, and 5 bins. |
| | *HSV Contrasts* [27] | 3 | Standard deviation of the HSV histograms. |
| | *Pleasure, Arousal, Dominance* [27] | 3 | Linear combination of HSV values according to visual perception statistics. |
| **Texture** | *GLCM* [27, 16] | 4 | Entropy, Energy, Contrast, and Homogeneity of the Gray-Level Co-Occurrence Matrix. |
| **Basic Quality** | *Contrast Balance* [34] | 1 | The $l_2$ distance between an original image and contrast-equalized image. |
| | *Exposure Balance* [34] | 1 | Absolute value of the skewness of luminance histogram. |
| | *JPEG Quality* [40] | 1 | No-reference quality estimation algorithm of [40]. |
| | *Sharpness* [34] | 1 | Sum of image pixels filtered with Sobel masks. |
| **Composition** | *Presence of Objects* [34] | 9 | Average saliency [17] on 9 image sub-regions. |
| | *Uniqueness* [34] | 1 | The $l_2$ distance between image spectrum and average spectrum of natural images. |
| | *Symmetry* [34] | 1 | Difference between HOG feature vectors of left / right image halves. |

**Table 1: Summary and description of the aesthetic features used for the supervised beauty model.**

**The stillness metric:** For each frame in the resulting subshots, we compute its *stillness* value as an inverse of the sum-squared pixel-wise frame difference value between two time-consecutive frames. This metric also represents the motion energy of a frame.

**Keyframe extraction:** We extract keyframes, one per subshot, by finding the most still frame from each subshot using the stillness metric. Figure 4 shows our approach consistently producing sharper key-frames compared to the conventional approach.

The stillness turns out to be an important aspect in determining the quality of thumbnails; we show this in Section 6 by analyzing the importance of the stillness in editorially chosen thumbnails.

The average processing time of our keyframe extraction was 0.27% relative to the video length, and the average number of keyframes was 2.1% relative to the total number of frames.

### 3.3 Thumbnail Selection

Finally, we select thumbnails from the extracted keyframes based on two main criteria, namely the *relevance* and the *attractiveness*.

**Frame relevance:** How much is the frame representative of the video content? To consider the relevance aspect, we perform a clustering analysis and measure the relevance of each keyframe based on the size of the clusters. Specifically, we cluster the keyframes using the k-means algorithm with the gap statistic method [39] in order to automatically find the optimal number of clusters. The gap statistic method does so by comparing the change in within-cluster dispersion with that expected under an appropriate reference null distribution. We varied the number of clusters from 5 to 10, and compared to 10 null reference distributions with each generated with 1,000 random samples. From the optimal clustering result, we generate thumbnail candidates, one per cluster, by selecting a frame with the highest aesthetic score, and rank them by their corresponding cluster size.

**Frame Attractiveness:** How much is the frame visually attractive? To consider the attractiveness aspect, we turn to computational aesthetics and introduce two versions of our system: unsupervised and supervised.

- The **unsupervised** approach is based on a simple heuristic that uses the *stillness* score; similar to Section 3.2, it selects a thumbnail from each cluster by finding the smallest frame difference value.
- The **supervised** approach is based on the recent computational aesthetics framework of [9, 30] that assesses the photographic attractiveness of images. We extract a set of visual features designed to capture various aesthetic properties, and train a random forest regression model [4] on a set of images annotated with subjective aesthetic scores. This model is used to assign a beauty score to each keyframe. Thumbnails are selected, one per cluster, by finding the highest scoring beauty score.

Next, we provide a detailed description of our supervised model for frame attractiveness scoring.

#### 3.3.1 Visual Aesthetic Features

Inspired by the work by Datta *et al.* [9], we designed a 52-dimensional vector of visual features that capture several important visual aesthetic properties (summarized in Table 1):

**Color:** The color distribution is one of the most closely related properties to visual aesthetics [9]. To evaluate the color composition and rendering, we compute: a *Contrast* metric based on luminance values; a set of Hue, Saturation, Brightness (HSV) statistics, including the *Average HSV* value for the whole image, the *Central Average HSV* value for the image central quadrant, the *HSV Color Histograms*, obtained by quantizing the values of HSV channels, the *Pleasure, Arousal, Dominance* statistics, and the *HSV Contrasts* [27], i.e. the standard deviation of the HSV histograms.

**Texture:** Inspired by previous work on image affect analysis [27], we describe the image texture by extracting the Haralick's features [16], namely the *Entropy, Energy, Homogeneity, Contrast* of the Gray-Level Co-occurrence Matrix (GLCM).

**Quality:** We include four basic image quality metrics to capture the level of degradation on video frames, typically caused by post-processing and compression: *Contrast Balance*, computed as the $l_2$ distance between the original frame and its contrast-normalized version after luminance histogram equalization; the *Exposure Balance*, computed as the skewness of the luminance histogram; a *JPEG Quality* metric based on the presence of JPEG blocking artifacts [40], and the global *Sharpness* of an image.

**Composition:** We describe the scene composition – the arrangement of objects in an image – by analyzing the distribution of the spectral saliency [17] across 3x3 image subdivisions, thus capturing by how much the image follows the golden ratio. This is commonly called the *Rule of Thirds* [9]. Similar to [33], we compute the *Symmetry* feature, which captures the left-right symmetry of an image, and the *Uniqueness* feature, which captures the originality of the image composition.

### 3.3.2 Supervised Attractiveness Scoring Engine

To learn a model able to score visual aesthetics, we use the AVA aesthetic database [30], which is the largest aesthetic database available in the literature. The dataset contains around 250,000 images with aesthetic, semantic, and style labels.

We designed a supervised framework to predict the aesthetic quality of video frames, using a random forest regression model [4]. We optimized the number of trees in our model via 5-fold cross validation, that is, we trained the model based on four splits (80% of the images) and tested it on the rest, repeating five times. We varied the number of trees from 1 to 200, and found that setting the number to 100 consistently provides good results. The average Mean Squared Error we obtained was 0.36 with a Multiple Correlation Coefficient (Pearson's correlation between predicted and real scores) of 0.49. We used this model throughout the paper to compute the aesthetic score of all the keyframes we extracted.

## 4. DATA COLLECTION

To validate the accuracy of our proposed framework, we consider a large dataset of videos with their corresponding thumbnails. The video dataset we used in our experiment comes from a random selection of 1,520 videos from Yahoo Screen (retrieved on February 2015). From the 1,520 videos, we discarded 402 videos where the provided thumbnail does not come from the actual video frames, resulting in 1,118 videos. In order to ensure a diverse sample set, our videos are roughly uniformly distributed among 18 different categories, as shown in Table 2.

### 4.1 Ground-Truth Thumbnails

Our dataset contains high-quality thumbnail images provided by a team of professional video editors; we consider the editorially chosen thumbnails as the ground-truth. This enables us to perform automatic evaluation of thumbnail selection without the hassle of obtaining labels via conventional means (e.g., crowdsourcing).

| Category | # Vid. | Mean Time | Total Time |
|---|---|---|---|
| Autos | 55 | 03:06 | 02:50:23 |
| Celebrity | 70 | 02:21 | 02:44:30 |
| Comedy | 70 | 03:45 | 04:23:00 |
| Cute & Inspiring | 75 | 01:50 | 02:17:19 |
| Fashion & Beauty | 76 | 02:35 | 03:16:38 |
| Finance | 52 | 03:14 | 02:47:59 |
| Food | 78 | 02:34 | 03:20:17 |
| Gaming | 46 | 04:01 | 03:04:43 |
| Health Fitness | 70 | 03:02 | 03:31:56 |
| International | 82 | 02:43 | 03:42:40 |
| Makers | 18 | 01:43 | 00:31:00 |
| Movie Trailers & Clips | 64 | 02:06 | 02:14:15 |
| News | 83 | 03:10 | 04:22:20 |
| Parenting | 13 | 02:12 | 00:28:31 |
| Sports | 53 | 02:24 | 02:07:17 |
| TV Highlights | 83 | 03:11 | 04:24:53 |
| Tech | 57 | 02:40 | 02:32:25 |
| Travel | 73 | 03:07 | 03:47:36 |
| *Total* | 1118 | 02:49 | 52:27:42 |

**Table 2: Descriptive statistics of our video dataset.**

To perform automatic evaluation, we need to know where exactly the thumbnail image appears in a video. We may search for a frame with the minimum sum-squared pixel-wise difference compared to the thumbnail image. Unfortunately, this method would not work because many thumbnails are edited from the original video frame via cut, zoom, color filters, and text overlays. In addition, not all thumbnails are actually selected from video frames. Our analysis revealed that about one-quarter of the videos have their thumbnails outsourced, e.g., from the Getty images.

In this work, we extract image features from the thumbnail image and video frames, compute an $l_2$ distance between each pair, and pick top 10 frames with the minimum error. We then manually select, among the top 10 frames, the one that looks most visually similar to the thumbnail image. We computed $l_2$ distances using HSV histograms and edge orientation/magnitude histograms, extracted over a three-level spatial pyramid. The selected frame's index is considered as the ground-truth. Through the manual verifying process, we filtered out 402 videos whose provided thumbnail does not appear in video, resulting in a total of 1,118 videos.

## 5. EXPERIMENTS

Can our system predict which thumbnails human editors would choose? We evaluate our system on a task of predicting editorially chosen thumbnails.

### 5.1 Methodology

We formulate the problem as a prediction task where the goal is to come up with a ranked list of "thumbnail candidate" frames $\mathcal{T} = \{F_1 \cdots F_n\}$. Different metrics have been proposed to evaluate the quality of ranked lists, such as precision at $k$ (P@$k$), mean reciprocal rank (MRR), and several variants of discounted cumulative gain (DCG). In this work, we consider a prediction is correct if *at least one* of the top-$k$ frames ($k \leq n$) matches with the ground-truth (i.e., editorial-chosen) frame $F^*$. With a slight abuse of terminology, we will also call this measure P@$k$ throughout the paper. We define P@$k$ of top-$k$ items from a ranked list $\mathcal{T}$ as

$$\text{P@}k(\mathcal{T}) = \begin{cases} 1, & \text{if } F^* \in \{F_1 \cdots F_k\} \\ 0, & \text{otherwise} \end{cases} \quad (4)$$

Below we report mean P@$k$ with $k = 1 \cdots 5$ across 1,118 videos.

Our problem setting reflects a practical thumbnail selection scenario: Depending on the situation, thumbnails might need to be generated either fully-automatically, by selecting a single best frame in the list ($k = 1$), or semi-automatically, by asking humans to choose a thumbnail from the top-$k$ frames according to our framework. In the latter case, the exact ordering among the frames becomes less important as long as they include at least one good thumbnail. We therefore assign the precision value 1.0 as long as one of the top-$k$ prediction is correct.

A video contains many near-duplicate frames; we consider a predicted frame as a match if it is visually "similar enough" to the ground-truth frame. To do this in an objective manner, similar to [20], we compute a similarity distance between the ground-truth frame and each of the video frames, and consider a frame as a match if its distance is smaller than some threshold $\theta$.

Specifically, we compute the SIFTflow [24] between each frame $F_i$ and the ground-truth $F^*$, warp $F_i$ into $\tilde{F}_i$ using the flow vector, and compute the sum-squared pixel-wise difference between $\tilde{F}_i$ and $F^*$. Based on empirical observation, we found setting the distance threshold value $\theta$ to 0.005 provides a reasonable bar that distinguishes near-duplicate of the ground-truth frame from irrelevant frames; we use this threshold throughout the experiment.

## 5.2 Models

We evaluated the two variants of our model (unsupervised and supervised) against six baseline models, described below.

All six baseline models skipped our frame filtering and keyframe extraction steps. Similar to our method described in Section 3.2, we used color and edge histograms over a two-level spatial pyramid as the feature representation for K-means centroid & stillness and group LASSO. The beauty rank model used 52-dimensional aesthetic features described in Section 3.3.1, and the CNN model used convolutional features extracted directly from images.

**Random**: We select and rank $k$ frames at random, without considering frame content.

**K-means centroid**: We perform k-means clustering over the entire frames of each video, using gap statistic [39] to estimate the optimal number of clusters, varied between 5 and 10. We then select top-$k$ frames from the $k$ largest clusters, choosing one per cluster with the closest $l_2$ distance to the centroid.

**K-means stillness**: Same as the above, except that we use the stillness metric to select a frame from each cluster.

**Group LASSO**: Similar to the sparse reconstruction method by Cong *et al.* [7], we select $k$ frames by solving group LASSO [28].
Let $\mathbf{X} \in \mathbb{R}^{d \times n}$ be a matrix of $n$ video frames, where each column is a $d$-dimensional feature vector extracted from each frame. We find sparse coefficients $\mathbf{A} \in \mathbb{R}^{n \times n}$ by solving the following convex optimization problem:

$$\arg\min_{\mathbf{A}} \|\mathbf{X} - \mathbf{X}\mathbf{A}\|_F^2 + \frac{\lambda}{2}\|\mathbf{A}\|_{2,1} \quad (5)$$

where $\|A\|_F^2 = \sum_{i,j} |a_{i,j}|^2$ and $\|A\|_{2,1} = \sum_i (\sum_j |a_{i,j}|^2)^{1/2}$. The first term measures the reconstruction error, the second is the row-wise sparsity inducing norm over $\mathbf{A}$, and $\lambda$ balances the relative importance between the two terms. After finding the optimal $\mathbf{A}$, we assigns the representativeness score $s_i$ to each frame as $s_i = \|\mathbf{A}_{i,:}\|_2$ and select top-$k$ highest scoring frames as thumbnail candidates. We experimented with $\lambda = \{0.1, 1.0, 10.0\}$.

**Beauty rank**: We use our supervised aesthetic scoring engine (see Section 3.3) to obtain top-$k$ thumbnails, examining every fifth frame of a video and selecting the top $k$ highest ranking frames.

| Models | Mean Precision@$k$ ($\theta = 0.005$) | | |
|---|---|---|---|
| | $k=1$ | $k=3$ | $k=5$ |
| Random | 0.0322‡ | 0.0778‡ | 0.1127‡ |
| K-means Centroid | 0.0483 | 0.1091† | 0.1610 |
| K-means Stillness | 0.0555 | 0.1118† | 0.1574† |
| G.-LASSO [7] | 0.0349‡ | 0.0823‡ | 0.1324‡ |
| Beauty Rank | 0.0125‡ | 0.0277‡ | 0.0367‡ |
| CNN [42] | 0.0411† | 0.0590‡ | 0.0689‡ |
| Ours Supervised | 0.0519 | 0.1190† | 0.1619 |
| Ours Unsupervised | **0.0653** | **0.1494** | **0.1896** |

Table 3: Mean P@$k$ averaged over 1,118 videos. The best performing result is bold-faced, the second best is underlined. We mark statistically significant differences relative to Ours Unsupervised: $p < 0.05$ (†) and $p < 0.001$ (‡).

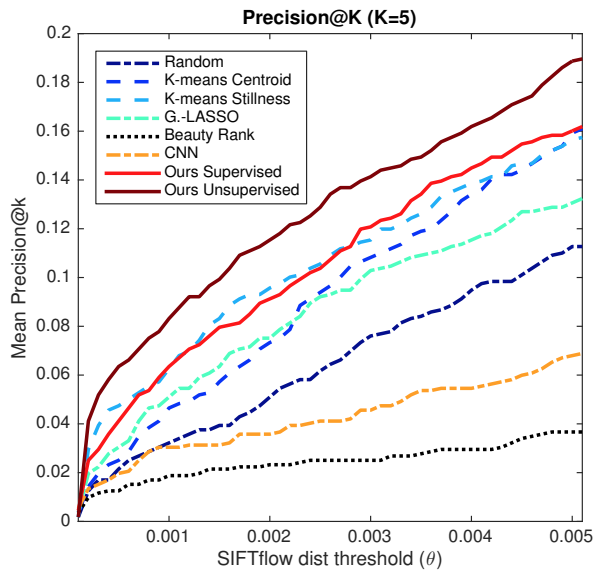

Figure 5: Mean P@5 across the SIFTflow distance threshold $\theta$.

**CNN**: We use a Convolutional Neural Network (CNN) to categorize a frame into either editorially chosen (positive) or randomly chosen (negative). Similar to Yang and Tsai [42], we trained our model using editorially chosen thumbnails as the positive examples and 5 random frames from each video as the negative examples.

The model was pre-trained on the ILSVRC2012 dataset [35] using the AlexNet network architecture [22]; we then fine-tuned the model on our dataset by changing the last layer to produce 2-dimensional output. We performed 5-fold cross validation and obtained the best model per split based on accuracy, after iterating 2,000 epochs. To obtain top-$k$ thumbnails, we examined every fifth frame of a video and picked the top $k$ highest ranking frames.

## 5.3 Results and Discussion

Table 3 shows the mean P@$k$ for $k \in \{1, 3, 5\}$, computed using the SIFTflow distance threshold $\theta = 0.005$ and averaged over all 1,118 videos. Figure 5 shows the full spectrum of the mean P@5 across different $\theta \in \{0, \cdots, 0.005\}$. Figure 6 shows the mean P@$k$ for $k = 5$ and $\theta = 0.005$, organized into different channels.

For the task of selecting a single best thumbnail ($k = 1$), an unsupervised version of our approach (Ours Unsupervised) outperformed all other approaches. The differences were statistically sig-nificant for the most cases (see Table 3). The supervised version of our approach performed the second best in $k > 1$. Figure 5 shows the relative performance differences are similar across difference values of the threshold $\theta$.

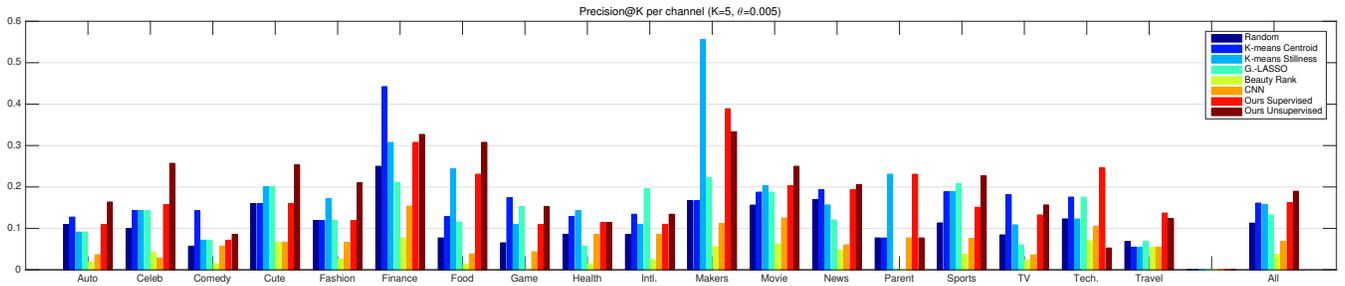

**Figure 6: Mean Precision@$k$ bar plots across different channels ($k = 5$, $\theta = 0.005$).**

A comparison between Ours Unsupervised and K-means Stillness shows the effectiveness of our frame filtering and keyframe extraction steps; the two models are identical except for those steps.

A comparison between both of our method and G.-LASSO [7], which was originally proposed as a video summarization method, suggests that conventional video summarization methods may not be directly applicable to thumbnail selection. G.-LASSO selects a set of frames that best summarizes the video content; this can be seen as optimizing for the relevance, similar to the two K-means methods. Our method considers both the relevance and the visual aesthetics; our results suggests the importance of considering both.

The poor performance of Beauty Rank and CNN, especially at $k > 1$, is due in part to the fact that it often selects near duplicate frames as the resulting thumbnails, because those frames will have similar aesthetic scores. Our framework prevents this from happening by removing near-duplicate frames via keyframe extraction, reducing redundancy in thumbnail candidate frames.

Finally, we show in Figure 9 some example pairs of thumbnails chosen by the editors and by our method (supervised version). To demonstrate the effectiveness of our method, we show the rank quantiles of visual features. A higher rank quantile means the frame has a higher feature value compared to other frames within the same video (Section 6 explains in detail how we compute the rank quantiles). A close look at each of the pairs and the rank quantile values of their visual features reveals that, even though our method sometimes select thumbnails different from the editors, the thumbnails tend to display high-quality visual characteristics.

In summary, on the task of predicting editorially chosen thumbnails, we showed that both versions our method performs favorably when compared to six baseline methods. This suggests that our method is able to select thumbnails that are visually similar to the ones that professional video editors would choose. In addition, qualitative comparisons on the resulting thumbnails show that, in some cases, thumbnails chosen by our method even possess higher-quality visual characteristics (e.g., brighter, sharper, and higher contrast thumbnails). This strongly suggests a need for a qualitative comparison of the produced thumbnails by human judgments. We leave this as an interesting future work.

## 6. QUALITATIVE ANALYSIS

In this section, we study *what makes a frame a good thumbnail* by investigating the influence of various visual aesthetic features on editorially chosen thumbnails. Specifically, we analyze the statistical relationship between thumbnail frames and non-thumbnail frames in terms of various visual features.

### 6.1 Methodology

**Visual features**: We consider a total of 53 visual features: the 52-dimensional aesthetic features of Table 1 and the *Stillness* score computed as an inverse of the sum-squared pixel-wise frame difference value between two consecutive frames. These features together encode various aspects of an image, including color, texture, basic quality, composition, and motion energy.

**Data preparation**: When comparing visual characteristics between thumbnail frames and non-thumbnail frames, we must consider the "nested structure" of the data: Frames are nested under a given video. Retaining this structure throughout the analysis is critical because different videos may have radically different characteristics, e.g., a baseball video can have more dynamic motion energy than a news video, and a black-and-white video will have a different shape of an HSV histogram compared to an RGB video. Consequently, "what makes a frame a good thumbnail" in one video may not apply to another. We must analyze visual characteristics of a thumbnail only within the context of a given video.

We prepare our data in the following way. For each video, and for each of the 53 visual features, we rank (by sorting the feature values) the editorially chosen thumbnail with respect to all other frames of the same video, excluding the thumbnail frame. We then represent the rank in the quantile range of [0,1]. For example, if the sharpness value of a thumbnail frame ranks at the 90% compared to all other frames of the video, its quantile is 0.9. This ranking method allows us to retain the nested structure of our data, representing each thumbnail only compared to other frames in the same video. To make the comparison more accurate, we remove near-duplicate thumbnail frames by applying the same technique we used in Section 5.1, which is based on the SIFTflow distance thresholding with $\theta = 0.005$. We compute the quantile of all features across 1,118 videos in this way, and obtain our data in the form a matrix of the size 1118-by-53.

**Statistical analysis**: We make the null hypothesis that there is no statistical relationship between each of the visual features and the selection of thumbnails. The rejection of such hypothesis will show the significance of the relationship. The standard way of testing hypotheses like this is using the chi-square goodness-of-fit test [31]. Specifically, we assume that each set of the 1,118 quantile values, computed using one of the 53 visual features, is uniformly distributed – this means that the editor's selection of a thumbnail was unrelated to the visual feature. If there is a statistically significant relationship between the visual feature and thumbnails, the distribution would fail to fit the uniform distribution; the chi-square test will reject the null hypothesis in such case.

In addition to the goodness-of-fit test, we analyze the patterns of visual features across 1,118 videos and report the means and standard deviations of quantile values across different visual features.

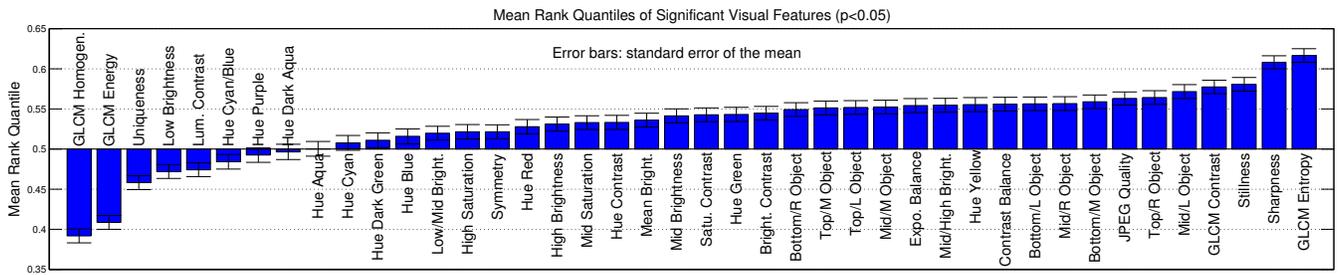

Figure 7: Mean rank quantiles of visual features whose influence on thumbnails were statistically significant ($p < .05$). A mean rank quantile above 0.5 (e.g., sharpness) means thumbnails tend to have a higher feature value among other frames of the same video.

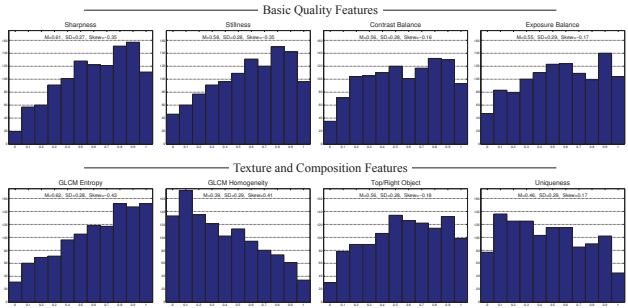

Figure 8: Histograms of rank quantiles for basic quality features (top) and texture and compositional features (bottom).

Such results will allow us to draw conclusions on the way the visual features influence the selection of thumbnails.

## 6.2 Results and Discussion

Figure 7 shows the summary of our results. It suggests that editorially chosen thumbnails tend to follow our general understanding of high-quality images in terms of the *Objective Quality* metrics (e.g., sharpness, brightness, contrast, and colorfulness). This explains the superior performance of the unsupervised version of our system; it is directly optimized for those metrics. On the other hand, the results also suggest that standard photographic beauty rules, such as the *Rule of Thirds*, may not be crucial for the selection of good thumbnails. Below, we discuss *"what makes a frame a good thumbnail"* by investigating each group of features from Table 1, and their relation with the selection of good thumbnails.

**Color:** The mean rank quantile values of many of the *Hue* values (see Figure 7) suggest that good thumbnails generally do not show a dominant color. Instead, their *Hue Contrast* (standard deviation of the Hue histograms, similar to the *colorfulness*) tend to be higher than other frames of a video. Moreover, we can see that dark frames are unlikely to be chosen as video thumbnails: The rank quantile of *Low Brightness* is below 0.5.

**Texture:** Figure 7 shows that editorial thumbnails have high entropy & contrast values, and low energy & homogeneity values on the GLCM. Figure 8 shows the distribution of rank quantiles for GLCM energy & homogeneity. This pattern suggests that, compared to other frames, thumbnails tend to have more complex compositions and non-smooth texture patterns, which means they tend to contain much more information than other frames. Note that this is different from the recent findings on photographic beauty [34], which showed that beautiful photos of faces tend to show smoother, homogeneous textures compared to lower quality images.

**Basic Quality:** We observe that basic quality metrics such as *Contrast Balance* and *Exposure Balance* tend to have higher values in editorially chosen thumbnails. We can also see that objective properties such as the frame *Stillness* and *Sharpness* serve as a good proxy to select good thumbnails; they are the second and the fourth most significant features according to the $\chi^2$ test. This suggests that editors specifically look for a high quality frame that stands out from other frames of a video, which may include those affected by camera motion and transition blur.

**Composition:** The results in Figure 7 on the *Presence of Objects* features (e.g., top/left object, bottom/right object, etc.) suggest that thumbnails do have a salient object, but in almost any position of the image. This implies that good thumbnails in general do not follow the *Rule of Thirds* photographic norm. It also suggests that non-thumbnail frames tend to have less presence of a salient object, driving their saliency distribution to the lower end. Similarly, our $\chi^2$ test revealed that subjective / artistic properties, such as *Uniqueness*, have less significant influence on editorial thumbnails compared to other features (it was the 28th significant feature). We note that this finding also contrasts with the photographic beauty [34].

## 7. CONCLUSION

We presented an automatic thumbnail selection system that exploits the relevance to video content and the visual attractiveness of frames. Our system identifies attractive thumbnails by analyzing various visual quality and aesthetic metrics, and performs a clustering analysis to determine the relevance to video content, thus making the resulting thumbnails more representative of the video.

We demonstrated the effectiveness of our system on the task of predicting thumbnails chosen by professional video editors, on a large-scale dataset of videos collected from Yahoo Screen. We showed empirically that our system performs favorably when compared to various other methods commonly used in the literature.

We also performed a qualitative analysis on *what makes a frame a good thumbnail* by analyzing the statistical relationship between thumbnail frames and non-thumbnail frames in terms of various visual features. We showed that the selection of good thumbnails is highly correlated with objective visual metrics, implying the possibility of building an automatic thumbnail selection system based on computational aesthetics.

Moving forward, there are many areas that could help improve our system. Previous research have used various forms of video metadata – such as video categories [38], video titles [37], search queries [25], and mouse tracking signals [45] – to improve performance on detecting thumbnails and highlights from videos. We are interested in exploring ways to leverage metadata. Our experimental results showed that performance varies across different video channels (see Figure 6), suggesting the potential for leverag-ing channel-specific information. Also, a video title is often carefully chosen to be maximally descriptive of its main topic [37]. Our work measured the relevance via visual frequency; measuring it via the relatedness to the title could provide better results.

Lastly, defining a reasonable standard evaluation protocol would be very critical in subjective tasks such as thumbnail selection; we believe our large-scale dataset and the design of experiments have contributed to make a step forward in that vein. Real-world evaluations, such as A/B testing by analyzing click-through-rates, could help us directly measure the quality of thumbnails and how they improve user engagement; it is the focus of our ongoing work.

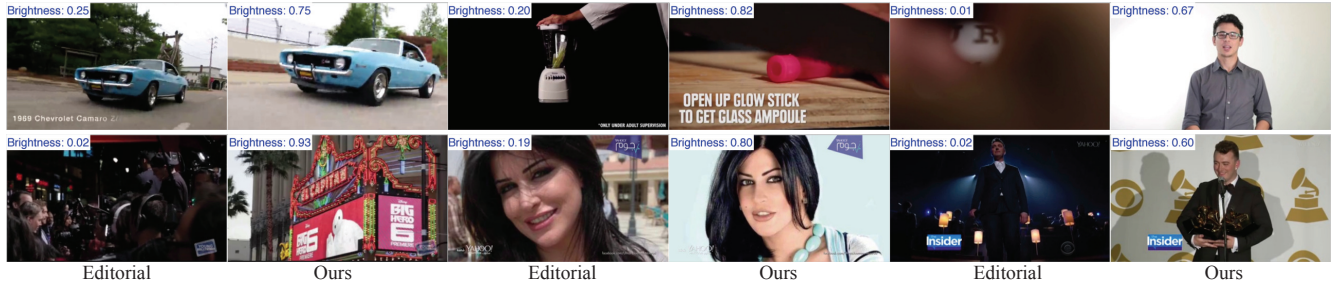
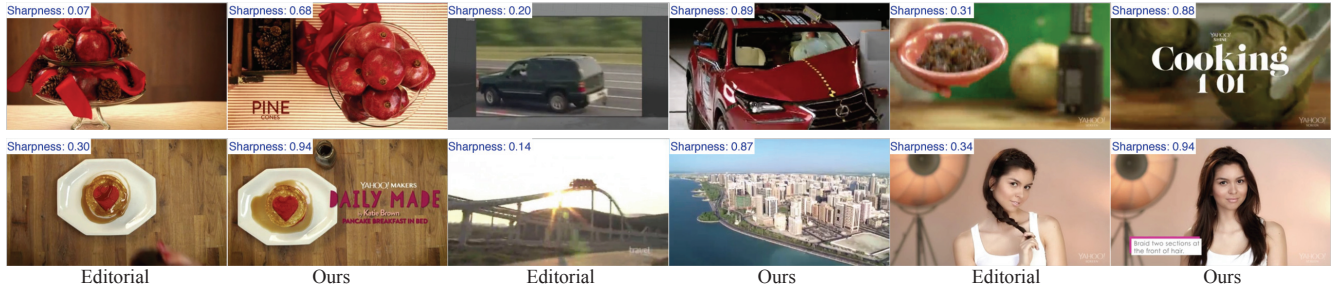
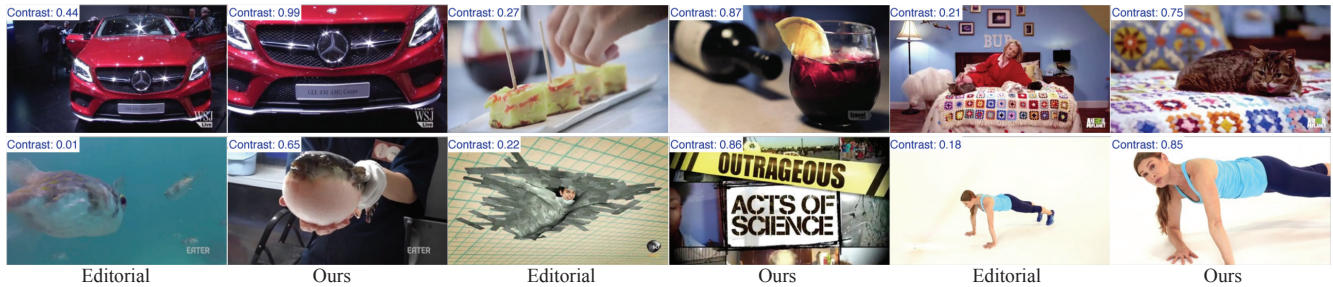

Figure 9: Qualitative comparisons on some example pairs of thumbnails chosen by professional video editors and by our method (supervised version). For each pair, we report the rank quantile values inside each image in order to support numerical comparison as well (a higher rank quantile means the frame has a higher feature value compared to other frames within the same video). While our method may select thumbnails different from the editorial ones, it is designed to select *beautiful* thumbnails. As a result, our thumbnails tend to have desirable visual characteristics, e.g., (a) brighter, (b) sharper, and/or (c) higher contrast frames.